\documentstyle[12pt,epsf]{article}
 
\newcommand{\msbar}{\overline{\mbox{\tiny MS}}}
\newcommand{\MSbar}{\overline{\mbox{\small MS}}}

\newcommand{\lbar}{\overline{\Lambda}}
\newcommand{\labar}{\overline{\Lambda}}

\newcommand{\beq}{\begin{equation}}
\newcommand{\eeq}{\end{equation}}
\newcommand{\be}{\begin{equation}}
\newcommand{\ee}{\end{equation}}
\newcommand{\beqn}{\begin{eqnarray}}
\newcommand{\eea}{\end{eqnarray}}
\newcommand{\bea}{\begin{eqnarray}}
\newcommand{\eeqn}{\end{eqnarray}}

\begin{document}
\pagestyle{empty}

\vspace{-0.6in}
\begin{flushright}
ROME prep. 1155/96
\end{flushright}
\vskip 2.0cm 
\centerline{\bf  THEORETICAL REVIEW OF $B$-PHYSICS}
\vskip 2.0cm
\centerline{\bf G.~Martinelli}
\vskip 1.0cm
\centerline{Dip. di Fisica, Universit\` a degli Studi di Roma
``La Sapienza'' and }
\centerline{INFN, Sezione di Roma, P.le A.Moro 2, 00185, Rome, Italy}
\vskip 5.8cm
\begin{abstract}

Weak decays and mixing of $B$-hadrons  play a special role in our
understanding of the physics of the Standard Model and beyond.
The measured
amplitudes, however, result from a complicated interplay of weak
and strong interaction effects.
Understanding  strong interaction dynamics, which becomes simpler
for heavy quarks, is thus a fundamental part of the game. In this review,
several theoretical aspects of $B$ dynamics  which are relevant for
current and future  experimental measurements are discussed.
\end{abstract}

\vskip 0.5 cm
\newpage
\pagestyle{plain}
\setcounter{page}{1}

\section{Introduction}
\label{sec:intro}
Hadronic weak decays are a unique source of information on the physics
of the Standard Model and beyond. Experimental measurements and theoretical
studies of these decays may provide us the key for understanding still unsolved
problems such as the reason for the existence of several fermion families,
the origin of  fermion masses and weak couplings and the mechanisms which
give rise to $CP$ violation. Through quantum-loop effects,
they are also a  window  which allows us to explore and constraint
new physics effects at high energy scales  still unaccessible to
particle accellerators,  such as those due to heavy supersymmetric particles.
 Unfortunately  the physical amplitudes result from a complicated interplay
of weak and strong interaction effects, which must be disentangled in
order to extract the values of the fundamental parameters.
\par In this framework,  $B$-hadron decays play a special role, 
since the $b$-quark belongs
to the heaviest, and less ``coupled", of the three known families and
 its partner, the top quark, has a mass which is  of the order
of the weak scale, $m_t  \sim G_F^{-1/2}$. This observation, for example,
 has led
to several speculations: that  spontaneous symmetry
breaking occurs via the Nambu-Jona Lasinio mechanism \cite{hill}, that 
quark masses are dynamically determined 
by radiative corrections \cite{zwirner} or that
the fundamental quark mass matrix, at large energy scales, is constrained
to have  very specific forms \cite{ross}. 
 Another aspect which is specific to
$B$-hadron decays is that the mass of the $b$-quark is much larger than the
scale of strong interactions, $m_b \gg \Lambda_{QCD}$.  In this 
situation,  QCD dynamics   becomes much simpler and in many cases, by
using the Heavy Quark Effective Theory (HQET) \cite{hqet},  we can obtain
accurate theoretical predictions to compare with the experimental measurements.
An example is given by the determination of the Cabibbo-Kobayashi-Maskawa 
(CKM) matrix
element $\vert V_{cb} \vert$ from exclusive ($B \to D^* l \nu_l$ and 
$B \to D l \nu_l$) and inclusive ($b \to X_c l \nu_l$) semi-leptonic decays.
 Since, however, the mass of the $b$-quark is large, but not infinite,
a related problem, to be discussed in the following,
 is that  of controlling
power corrections, i.e. corrections suppressed as inverse powers of the heavy
quark mass. 
 \par In this article,  I review  several theoretical issues in 
the physics of $B$ decays which  are relevant for  current or future
experimental measurements. A
special attention is paid to the discussion of the systematic errors present
in  the different theoretical approaches, and of the possible progresses
that we may expect in the near future. In preparing this talk, it 
has been unavoidable to make a
selection of  subjects.  Among those
 for which   new theoretical studies or new experimental  measurements appeared
this year, I have chosen those topics which look to me particularly
interesting, and/or for  which I have enough compentence  to express a  personal
point of view.   For lack of space,  I am unable to cover here radiative
decays and supersymmetric effects in FCNC processes. A discussion
on these points can be found in the talks by A.~Ali and A.~Masiero at this
workshop.
For further  information on  other subjects related to $B$-physics,
 not covered here, the reader can also refer to
some recent  review papers by  A. Ali \cite{alirev},
 A. Buras \cite{burasrev} and M. Neubert  \cite{nrev}.
\par This article  is organized as follows.  In sec.~\ref{sec:lepto},  a brief
overview  of  leptonic decays, $B^0$--$\bar B^0$ mixing and CP violation is
given; in sec.~\ref{sec:exsemilepto},  $B$-meson 
 exclusive semi-leptonic decays, and the extraction
 of $\vert V_{cb} \vert$ and $\vert V_{ub} \vert$, are
discussed; some aspects of exclusive non-leptonic decays are reviewed
in sec.~\ref{sec:exnonlepto}; power corrections
in the HQET are discussed in sec.~\ref{sec:power}; 
some considerations on inclusive semi-leptonic decays can be found
in sec.~\ref{sec:insemilepto};
the present status, and
unsolved problems, for $B$-hadron lifetimes will be summarized in 
sec.~\ref{sec:innonlepto}, followed by a brief conclusion, which contains
also an outlook on possible improvements in the accuracy
of  the theoretical  predictions, sec.~\ref{sec:conclu}.
\section{ $f_B$, $B^0$--$\bar B^0$ mixing and CP violation}
\label{sec:lepto}
The leptonic decay width of the $B$ meson is given by
 \beq \Gamma(B^+ \to \tau^+ \nu_\tau)=\frac{G_F^2 \vert V_{ub}\vert^2}{8 \pi}
M_B \left( 1 - \frac{M_\tau^2}{M_B^2}\right)^2 M_\tau^2 f_B^2 \label{eq:lepto}
\ .\eeq
A precise knowledge of the leptonic decay constant $f_B$, analogous to $f_\pi$
in $\pi \to \mu \nu_\mu$ decays~\footnote{ I use the normalization convention
in which $f_\pi=132$ MeV.}, would allow an accurate extraction of the CKM matrix
element $\vert V_{ub}\vert$, which is actually known with
a relative error of about $25 \%$, see below. One of the most
precise determinations of $f_B$ is presently given by lattice QCD calculations
and it is shown in table~\ref{tab:lr}, together with other results
for several physical quantities of interest for 
$B$-physics~\cite{martihf95}--\cite{sac}. The error on $f_B$
reported in the table comes mainly from the extrapolation of the lattice results
to the continuum  limit and does not includes the systematic error due to
the quenched approximation, which has been
estimated to be of the order of $10$--$30 \%$~\cite{sharpe}. A  global theoretical
error of $25$  MeV, corresponding to a relative error of about $14 \%$,
would provide us with an   accurate determination of $\vert V_{ub}\vert$,
through a measurement totally independent from the semi-leptonic channel,
and with  experimental and theoretical systematic effects different from
that case. Even though the rate (which goes as the squared lepton
mass) is  much smaller,
 the $B^+ \to \mu^+ \nu_\mu$ leptonic decay channel is also quite
interesting, because the much smaller number
of events  could  be compensated by a much
higher efficiency in detecting the final state charged lepton.
\begin{table}
 \centering 
\begin{tabular} {||c|c|c|c||}\hline  
$f_D$ (MeV) & $f_{D_s}$ (MeV)   & $ f_B$ (MeV)  & $f_{B_s}/f_B$ \\ \hline
$205 \pm 15$ & $235 \pm 15$ & $175 \pm 25$ & $1.15 \pm 0.05$ \\ \hline
$\hat B_{B_d}$ extrapolated & $\hat B_{B_d}$  HQET &
 $\hat B_{B_d}/\hat B_{B_s}$ & $\xi_d$  (MeV)\\ \hline 
$1.40 \pm 0.10$ & $1.08 \pm 0.06 \pm 0.08$ &
$1.01 \pm 0.01$ & $207 \pm 30 $ \\
\hline \end{tabular} \label{tab:lr}
\caption{ \it{Lattice predictions for several
quantities of interest in heavy flavour physics. A detailed explanation
of the methods used to compute these quantities, and of the evaluation
of the systematic errors can be found in 
\protect\cite{martihf95}--\protect\cite{sac}.
The numbers given here are my personal compilation of lattice
results obtained from the numbers reported
in refs.~\protect\cite{martihf95}--\protect\cite{sac} and the 
results which can be found in  ref.~\protect\cite{vince}.}}
\end{table}
\par  $f_B$ also enters in  phenomenological analyses 
of the $B^0$--$\bar B^0$ mixing amplitudes, together with the so called
renormalization group invariant $B$-parameter $\hat B_B$. The 
square of the mixing parameter  $\xi = f_B\sqrt{ \hat B_B}$
is in fact related to the matrix element of the renormalized
$\Delta B=2$  Hamiltonian \cite{burasdb2}.  All theoretical
calculations of
 $\hat B_B$ tend to give values very close to one for both the $B^0_d$
and the $B^0_s$ mesons, see table~\ref{tab:lr} and ref.~\cite{narison} for
QCD sum rules. Thus the strength of the mixing is essentially
regulated by the meson decay constant.  Taking into account the
theoretical uncertainties, and the experimental errors on the relevant CKM
matrix elements and on the $B$ lifetime, one predicts
$\Delta M_{B_d}= 0.55 \pm 0.13$ ps~\cite{ciuup}, to be compared with the 
experimental world average  $\Delta M_{B_d}= 0.464 \pm 0.018$ ps. \par
The value of $f_B$  is also important for CP violation in
$B$ decays.
It was realized long time ago
\cite{reina} that a
value of $\xi  \sim 200$ MeV, combined with a large value of the top mass,
(in the following
I will use $m_t^{\overline{MS}}(m_t^{\overline{MS}})= 
167 \pm 8$ GeV) leads to a 
large value  of $\sin 2\beta$, the parameter
which  controls CP violation in $B \to J/\psi K_s$ decays
\beq {\cal A} =\frac{N(B^0_d \to J/\psi K_s)(t)- N(\bar B^0_d \to J/\psi K_s)
(t)}{N(B^0_d \to J/\psi K_s)(t)+ N(\bar B^0_d \to J/\psi K_s)
(t)}=  \sin 2 \beta \sin \Delta M_{B_d} t \ . \eeq
\begin{figure}   
\begin{center}
\epsfxsize=1.08\textwidth
\leavevmode\epsffile{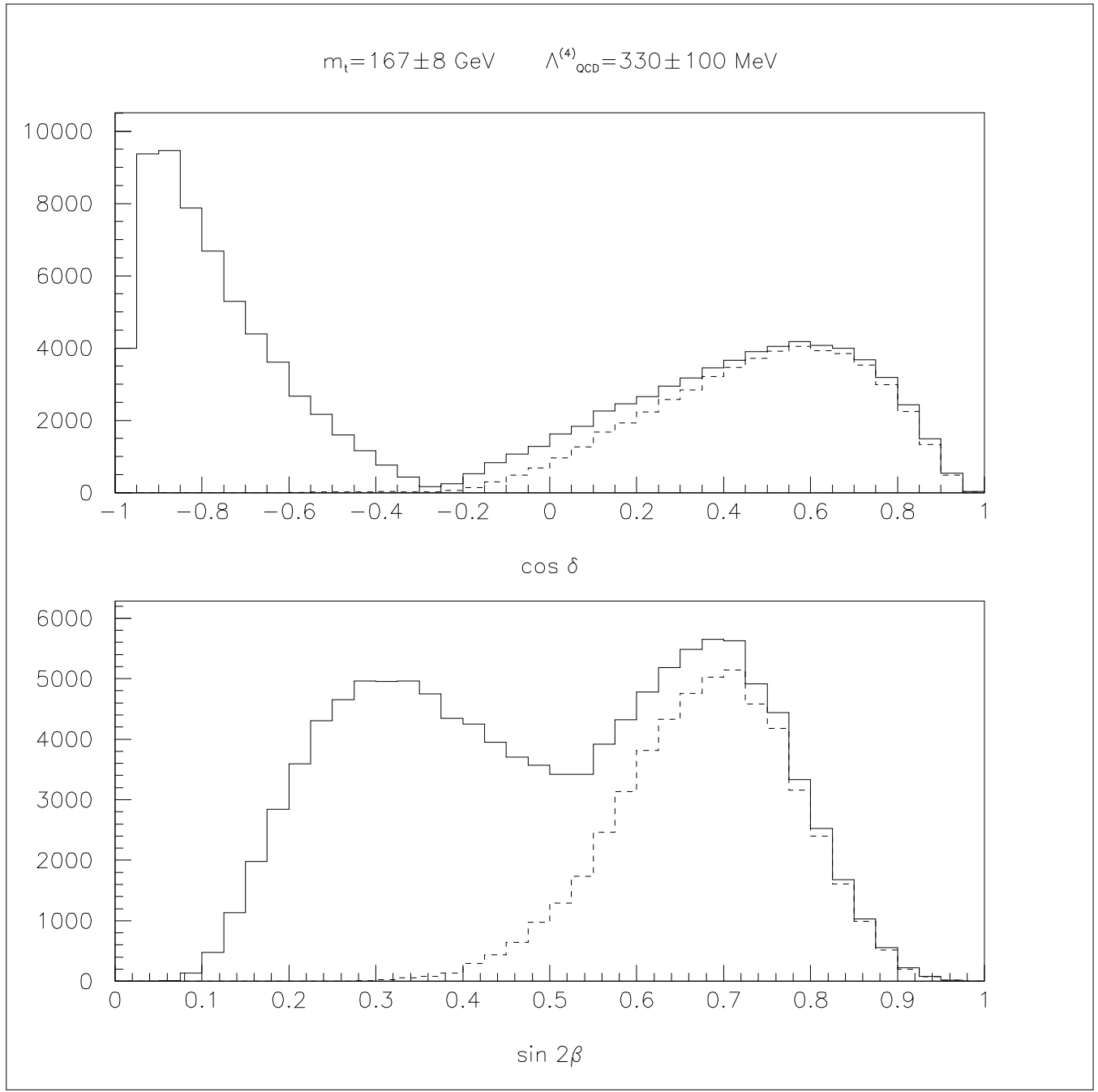}
\caption[]{\it{ Distributions of values for $\cos \delta$ and 
$\sin 2 \beta$,
for $m_t=(168 \pm 8)$ GeV.
The values of the parameters are those used in the second
of refs.~\protect\cite{reina}. The solid histograms are obtained without using
the information coming from  the experimental determination
of $\Delta M_{B_d}$. The dashed ones uses 
$\Delta M_{B_d}=0.464 \pm 0.018$, with the condition
$177 $ MeV $\le \xi_d=f_{B_d}\sqrt{\hat B_{B_d}} \le 237$ MeV.}}
\label{fig:sb}
\end{center}
\end{figure}
This is shown in fig.~\ref{fig:sb} where  the distribution of acceptable values
of  $\sin 2\beta$, obtained
from a combined analysis of the kaon CP violation
parameter $\epsilon$ and of the $B^0_d$--$\bar B^0_d$ mixing rate 
$\Delta M_{B_d}$ is shown
\cite{ciuup}. The histogram refers to the distribution of values of $\sin
2\beta$ obtained by varying within their errors the values 
of the different  experimental and theoretical
quantities which enter in the theoretical predictions. The dashed histogram
is obtained as the previous one, by imposing the further condition that
$\xi = 207 \pm 30$ MeV.  From the dashed distribution shown 
in the figure, one gets  $\sin 2\beta=0.68 \pm 0.10$, a value quite reassuring
for future experiments at the $B$-factories.
\par The reliability of the lattice estimates of the meson decay constants
is  demonstrated by the fact that $f_{D_s}$, which was predicted long before
the first experimental measurement \cite{latt88}, agrees with the present
experimental world average from $D_s \to  \mu \nu_\mu$
decays, $f_{D_s}=241 \pm 21 \pm 30$ MeV
($f_{D_s} \sim 260$ MeV if one includes the L3 measurement from
$D_s \to  \tau \nu_\tau$) \cite{richman}.  Further, indirect information
can be extracted from the experimental lower bound on the $B^0_s$--$\bar B^0_s$
mixing amplitude.
Taking  $\Delta M_{B_s} \ge  9.2$ ps  from experiments and
$f_{B_s}/f_B=1.15 \pm 0.05
^{+0.15}$ from  theory, where the last error is an estimate 
of the error coming from the 
quenched approximation~\footnote{ In the ratio of the decay constants
many systematic errors cancel.}, one obtains  $f_B \ge 160$  MeV in
good agreement with the direct theoretical prediction~\cite{ciuup}.
\par I want to conclude this section with a comment on the ``heavyness"
of the charm quark. The HQET is not a model. It is QCD systematically
expanded in powers of a small parameter
 $\mu_{QCD}/m_Q$, where $m_Q$ is the quark mass and $\mu_{QCD}$ stands for a
scale of the order of the strong interaction scale. Thus if 
$\mu_{QCD}=\Lambda_{QCD} \sim 300$ MeV, since $m_{charm} \sim
1.2$--$1.5$ GeV,  the expansion
parameter  is a small number even in the charm case. If, instead,  
$\mu_{QCD}=2 \pi f_D \sim 1.2$ GeV or of the order of the nucleon mass, the
expansion parameter is of order one and  the heavy quark expansion is
expected to fail. Lattice studies of the behaviour of the meson decay constant
$f_M$ as a function of the inverse meson mass \cite{martihf95}--\cite{sac}
 have shown 
that the expansion parameter is, at least in this case, of the order of 1 GeV.
These results have been confirmed by calculations using  
QCD sum rules~\cite{narison}. 
Such a large value of
$\mu_{QCD}$ implies that power corrections are   $\sim 80\%$--$100 \%$ for the
charmed meson decay constant, and  $\sim 25 \%$--$30 \%$ 
in the  $B$-meson case. Corrections
of the same order are also expected, and were indeed found in lattice
calculations, for the semi-leptonic  form factors relevant in
$B \to \pi$ and $B \to \rho$ decays~\cite{flynn}. A special, and probably
unique, case where the
application of the HQET to the charm is succesful, because of  Luke's
theorem \cite{luke}, is represented by the semi-leptonic $B \to D$ and
$D^*$ decays  which are discussed in sec. \ref{sec:exsemilepto}. In
all the other cases, as in the
discussion of the charmed hadron lifetimes of 
sec.~\ref{sec:innonlepto}, there is strong
experimental evidence that the charm is not heavy enough  to make  HQET
applicable.
\section{Exclusive semi-leptonic decays}
\label{sec:exsemilepto}
In this section I will discuss some theoretical issues related to exclusive
heavy-to-heavy and heavy-to-light semi-leptonic decays. 
\subsection{$V_{cb}$ from $B \to D^* l \nu_l$}
\par In heavy-to-heavy decays, a golden plated determination of
$\vert V_{cb} \vert$ is given by the comparison of the differential decay
rate for the process $B \to D^* l \nu_l$ to its theoretical 
expression~\footnote{ For lack of space I will not discuss the 
process $B \to D l \nu_l$ for which both measurements, and theoretical
predictions, exist.} 
\beqn \frac{d\Gamma}{dw}&=& \frac{G_F^2}{48 \pi^3} (M_B -M_{D^*})^2
M_{D^*}^3\sqrt{w^2-1} (w+1)^2 \nonumber \\ &\times& \left(
1+ \frac{4w}{w+1} \frac{M_B^2- 2 w M_B M_{D^*} +M_{D^*}^2}{(M_B-M_{D^*})^2}
\right)\ \vert V_{cb} \vert^2 {\cal F}^2(w) \  ,\label{eq:slg} \eeqn
where the variable $w$, related to the momentum trasfer $q$, is given by
\beq w = v_B \cdot v_{D^*}= \frac{E_{D^*}}{M_{D^*}}= \frac{
M_B^2 + M_{D^*}^2-q^2}{2 M_B M_{D^*}} \ .  \label{eq:iw}\eeq
In the infinite mass limit, up to logarithmic 
corrections,  ${\cal F}(w)$ tends to the Isgur-Wise function $\xi(w)$. 
When both $m_b$ and $m_{charm}$ become very large, the normalization of
$\xi(w)$  at zero recoil is fixed by the conservation of the weak
current, i.e. $\xi(w=1)=1$.  In these limits then, up to the trivial kinematical
 factors
appearing in  eq.~(\ref{eq:iw}),  a model independent extraction of 
$\vert V_{cb} \vert$ is possible. 
Two elements are important for an accurate
determination of $\vert V_{cb} \vert$, however: the control of the corrections
due to different  effects at zero recoil, and the extrapolation of the 
rate to $w=1$. \par i) The normalization
of ${\cal F}(w=1) $ is modified by short-distance strong interactions effects,
which can in principle be computed in perturbation theory, and by 
non-perturbative $1/m_Q$
corrections (starting at second order by virtue of  Luke's theorem
\cite{luke}). We have then 
\beq {\cal F}(w=1)= \eta_A \Bigl( 1 + c_2 \frac{\mu^2_{QCD}}{m^2_Q} + \dots
\Bigr) = \eta_A \Bigl( 1 + \delta_{1/m^2}  \Bigr) \ . \eeq
A recent two-loop calculation  has reduced the uncertainty coming
from the perturbative corrections to less than $1  \%$, giving
$\eta_A=0.960 \pm 0.007$ \cite{cer}.  The evaluation of $\delta_{1/m^2}$
is more problematic, since this is a non-perturbative quantity. 
Two different approaches were followed in the past. On the one hand, a
classification of all the $1/m_Q^2$ local operators of the HQET contributing to
$\delta_{1/m^2}$ was made, and their matrix elements estimated in a model
dependent way \cite{ndeltino}. These estimates gave $-\delta_{1/m^2}
=(3 \pm 2)\%$. On the other hand,
zero-recoil sum rules, which are in principle model-independent, gave a much
larger correction corresponding to $-\delta_{1/m^2}=5.5 \%$--$8 \%$ 
\cite{rdeltino,ural}. After some controversy, M. Neubert, by using a ``hybrid"
method, found a ``Solomonic" result, which is ``not-in-disagreement"
with both the values given above, namely $-\delta_{1/m^2}=(5.5\pm 2.5) \%$
\cite{hybrid}. This prediction is currently adopted in all the estimates
of the theoretical error on $\vert V_{cb} \vert$ extracted from the
experimental measurements of the exclusive semi-leptonic decays. \par 
A recent study of the accuracy of power corrections in effective theories,
illustrated by several explicit examples, showed that the predictions 
from the zero recoil sum rules, though model independent, are potentially 
subject to large corrections coming from higher order perturbative terms
which have not been computed yet \cite{msreno}. The same arguments apply
to the calculations of refs.~\cite{ndeltino,hybrid}. For this reason,
and given the high accuracy reached by the experimental measurements, I believe
that a further effort is needed to try to reduce, and to evaluate more
realistically, the error on $\delta_{1/m^2}$. This is very difficult because
the deviations due to $\delta_{1/m^2}$ are rather small, and no theoretical
method, including lattice simulations, can control the 
uncertanties to a sufficient degree of accuracy.
\par ii) Since the rate vanishes at $w=1$ (see eq.~(\ref{eq:slg})) in
order to extract ${\cal F}(w=1)$ an extrapolation of the data is 
necessary. This extrapolation is usually done by expanding the form factor
around $w=1$
\beq {\cal F}(w)= {\cal F}(1) \Bigl( 1- \hat \rho^2 (w-1) + \hat c (w-1)^2
+ \dots \Bigr) . \label{slopes} \eeq
The slope $\hat \rho$ and the curvature $\hat c$ can be fitted directly
from the experimental measurements of the differential rate~\footnote{ Following
ref.~\cite{nrev}, I use here $\hat \rho$ to denote the physical
slope, in order to distinguish it from $\rho$ which refers to the
slope of the Isgur-Wise function $\xi$.} 
\cite{artuso,gibbons}.  It is, however, helpful for reducing
the uncertainty of the extrapolation  to have  theoretical estimates 
of these parameters, for which several bounds, and calculations, obtained
using different techniques, exist. 
The spread of the theoretical predictions for $\hat \rho$ is,
unfortunately,  still rather large \cite{martihf95}, 
while for $\hat c$ there are
only analyticity bounds which relate its value to that of $\hat \rho$
\cite{ncap}.  In analogy to what is done experimentally,
 $\hat \rho$ has been  extracted  from lattice calculations
by fitting the  form factor computed in numerical simulations
to eq.~(\ref{slopes}). 
This fitting  procedure, however,  introduces  a rather large uncertainty
in the final result~\cite{martihf95}. 
I want to mention that a method to obtain directly
the $\hat \rho$ and $\hat c$, without any fitting procedure, or perturbative
calculation, has been proposed in ref.~\cite{agli}, but only a preliminary
feasibility study of this technique has been done to date \cite{lel}.
Contrary to the case of $\delta_{1/m^2}$, it is likely that more accurate
theoretical results for $\hat \rho$ will be obtained in the near future.
\par
\begin{figure}   
\begin{center}
\epsfxsize=1.08\textwidth
\leavevmode\epsffile{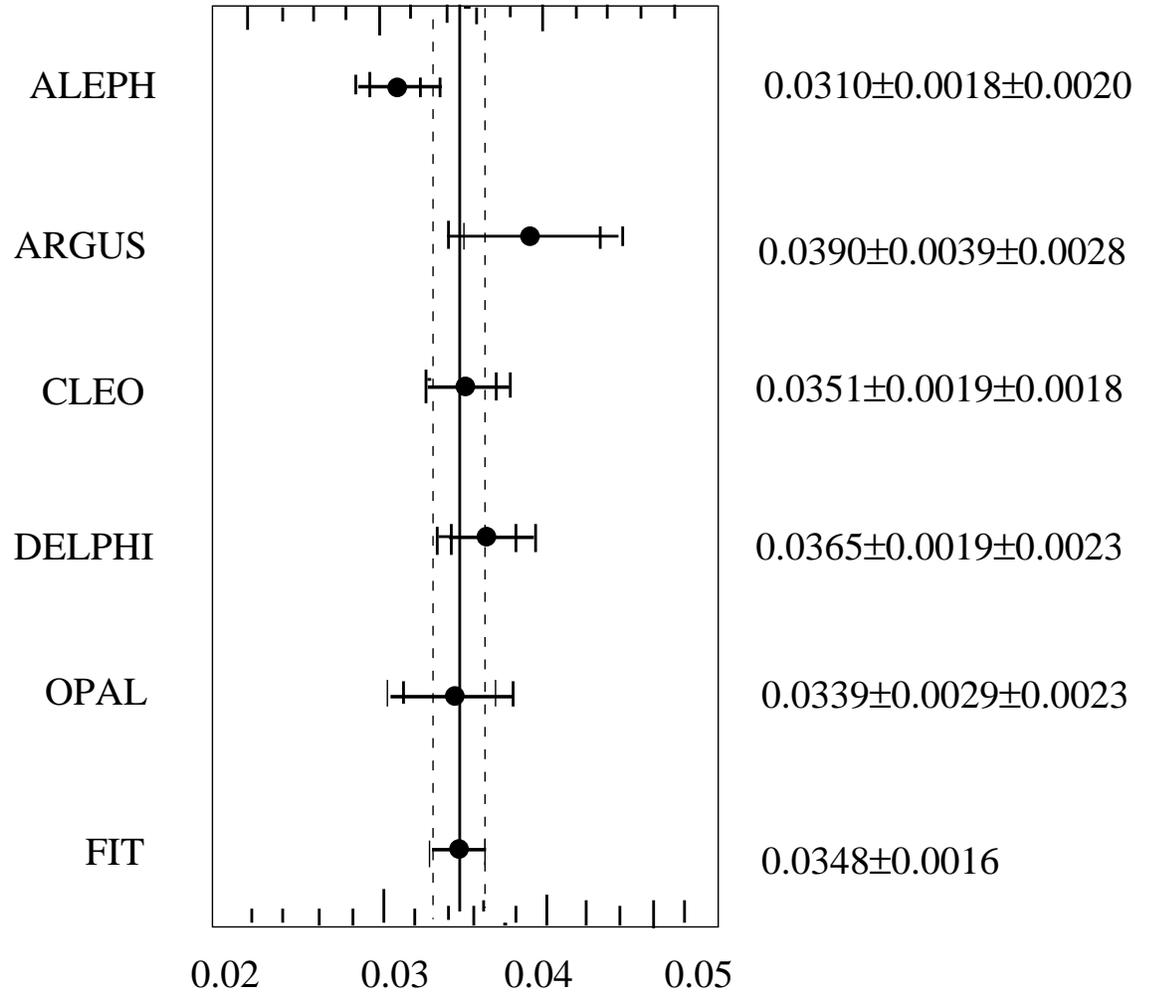}
\caption[]{\it{Determinations of $\vert V_{cb} \vert {\cal F}(w=1)$
from different experimental measurements. The average is also given.}}
\label{fig:vcb}
\end{center}
\end{figure}

 As shown in fig.~\ref{fig:vcb}, the values of ${\cal F}(w=1) \vert
V_{cb} \vert$ extracted by the different 
experiments are quite consistent. 
This is true
in spite of the fact that the fitted values of $\hat \rho$ are somehow 
different \cite{gibbons}. From the world average of  ${\cal F}(w=1) \vert
V_{cb} \vert$, by assuming from the  theory ${\cal F}(w=1)=0.91 
\pm 0.03$ one obtains,
\beq \vert V_{cb}\vert = 0.0382 \pm 0.0018 \pm 0.0012 \ , 
\label{eq:vcbex}\eeq
where the last error comes from the estimated theoretical uncertainty.
\subsection{ $V_{ub}$ from $B \to \rho l \nu_l$
and $B \to \pi l \nu_l$}
For heavy-to-light semi-leptonic decays,  HQET is less useful since
the normalization of the current is not fixed by symmetry. Still scaling laws
can be established for $q^2 \sim q^2_{max}$. For example, 
up to logarithmic corrections, one finds
\cite{scaff,abgla}
\beq f^+_P(\vec p_\pi)= A M_P^{1/2} \left(1 + \frac{\delta}{M_P} +\dots \right)
\label{eq:scal} \ ,\eeq
where $f^+_B(\vec p_\pi)$ is the form factor relevant in  $B \to \pi$ decays.
$P$ denotes the heavy pseudoscalar meson, $M_P$ its mass, $\vec p_\pi$
the pion momentum and $A$ and $\delta$  are non perturbative quantities.
The scaling law in eq.~(\ref{eq:scal}), and others  which can be derived
for the form factors relevant in $B \to \rho l \nu_l$ or 
$B \to K^* \gamma$ decays, are valid in the heavy meson rest frame, for
small values of the light meson three-momentum, i.e. for $\vec p_\pi
\ll M_P$.  Scaling laws at $q^2 \sim 0$ can also be derived by computing the
form factors using light-cone wave functions \cite{braunsca}. For example
one finds $A_1(q^2=0) \sim M_P^{-3/2}$, where $A_1$ is the axial form
factor entering in $B \to \rho l \nu_l$ decays. Unfortunately,
corrections to the asymptotic scaling behaviour were found
to be large both in lattice calculations 
\cite{flynn,abgla}
and using QCD sum rules \cite{braunsca}, cf. also the discussion on power
corrections in sec.~\ref{sec:lepto}. A further constraint 
which can be used in theoretical predictions is  that coming from
the equality of some of the form factors in the helicity basis  at $q^2=0$,
e.g.  $f^+(q^2=0)=f^0(q^2=0)$, and from analyticity  \cite{lello}.
The above information is particularly useful for lattice calculations,
where only a modest range of the physical $q^2$ is presently accessible.
\par
\begin{figure}   
\begin{center}
\epsfxsize=0.6\textwidth
\leavevmode\epsffile{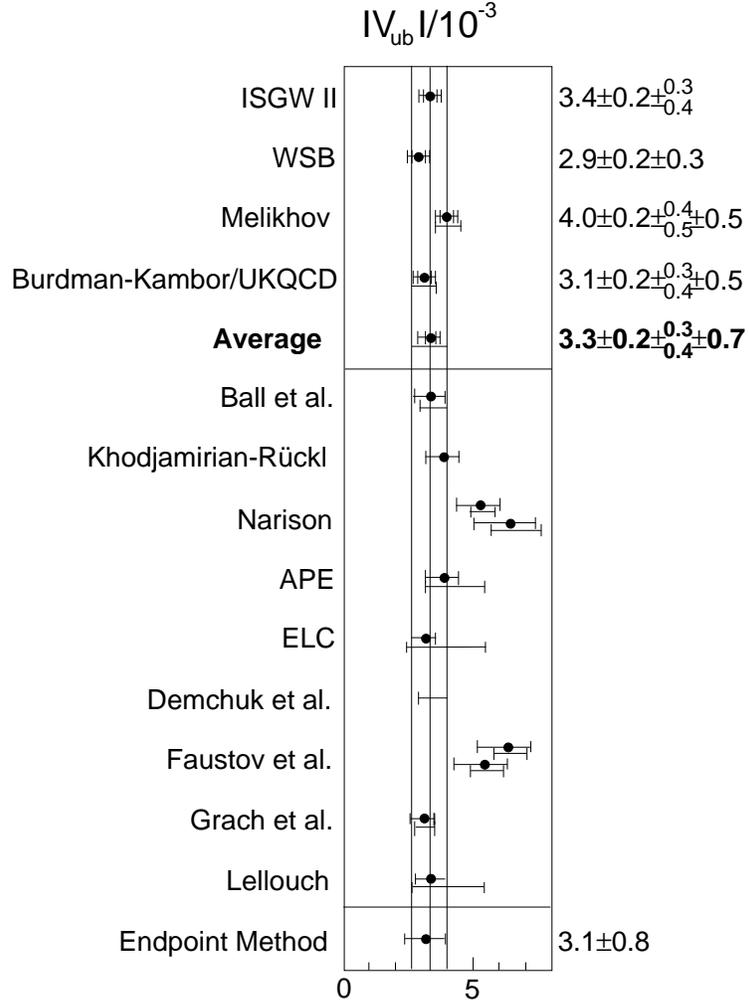}
\caption[]{\it{Experimental determinations of $\vert V_{ub} \vert$
from exclusive $B \to \pi$ and $B \to \rho$ semi-leptonic decays.
The results depend on the theoretical model used in the experimental
analysis. Only the models above the orizontal line have been used to extract
$\vert V_{ub} \vert$ from the data. The value of $\vert V_{ub} \vert$
from inclusive decays is also given at the bottom of the figure.}}
\label{fig:vub}
\end{center}
\end{figure}

 Several theoretical methods have been used to predict the form
factors for $B \to \pi$ and $B \to \rho$ decays. Besides the lattice 
calculations \cite{flynn} and the QCD sum rules
\cite{braunsca} mentioned above, several quark model calculations
have also appeared in the literature. A complete list
of theoretical calculations can be found
in refs.~\cite{richman,artuso,gibbons}. It is interesting that, with
the increasing experimental precision, it is possible
to discriminate the ``good" from the  ``bad" models by confronting
the ratio of widths $\Gamma(B \to \rho)/\Gamma(B \to \pi)$, in  which
the unknown $\vert V_{ub} \vert$ cancels, to the
theoretical predictions  \cite{richman,artuso,gibbons}. 
A compilation of results,
obtained by the CLEO collaboration using different theoretical models
is shown in fig.~\ref{fig:vub}.  
In the extraction of $\vert V_{ub} \vert$, 
 some models, the results of which are given in the
figure above the horizontal line,
have also been used in the Monte Carlo analysis of the data, in order 
to generate  events to be confronted to the experimental measurements. The
corresponding values for $\vert V_{ub} \vert$ are only given in these cases.
The last value, denoted as ``Endpoint method" refers, for comparison, to
the  result obtained from inclusive charmless decays, to be discussed
in the following.  From the figure, we see that the dominant error is
the theoretical one. This is true
 in spite of the encouraging progresses made by  the
theoretical calculations in the last few years. These progresses
are evident if one compares the results of
fig.~\ref{fig:vub} to those available at the time of the 5th  Symposium
on Heavy Flavour Physics, held in Montreal in 1993 \cite{hf93}. 
A further improvement
of the present situation is likely to occur in the near future, particularly
from lattice calculations.
\section{ Exclusive non-leptonic decays}
\label{sec:exnonlepto}
In this section, I briefly discuss a few items connected to exclusive
two-body non leptonic decays.
\vskip 0.2cm \underline{Factorization}: Factorization is an interesting
phenomenological idea which allows us to predict
non-leptonic amplitudes 
in terms of   semi-leptonic form factors and of  vector and pseudoscalar
decay constants.  In a way, it corresponds to an elaborated version of the
vacuum saturation approximation. 
Factorization   was originally  used to estimate kaon decay
amplitudes \cite{russia}. In the eighties, it  was applied
by Stech and collaborators in analyzing $D$-meson decays \cite{stech}
and  its explanation in terms of the $1/N$ expansion, where $N$
is the number of colours, was  subsequently proposed \cite{rucl}.
More recently, it has been suggested that factorization holds
in the heavy quark limit, for
some  energetic decay channels, because the ultrarelativistic
light mesons in the final state do not have time to exchange gluons 
\cite{bj,gr}. In ref.~\cite{agli2} it was shown, however, that
the demonstration of factorization  given in  \cite{gr}   has some problems. 
It may be still true  that factorization becomes valid  in the infinite 
mass limit, but a convincing proof of this intuitive physical
idea is still missing. 
Several observations are necessary at this point. \par i)
Factorization does not work for charm meson decays, where, moreover,
large final-state interaction effects are measured (final-state interactions
are zero if amplitudes factorize). This may be due either to a failure
of factorization, or to the fact that, as observed above, the charm
 quark is not heavy enough.  Indeed factorization, combined with
the assumption that final-state interactions are dominated by
nearby resonances, has been succesfully applied to phenomenological
studies of charmed meson decays, for example in ref.~\cite{lusi}.
\par ii) The effective Hamiltonian which controls weak charm decays
has the form
\beq {\cal H} = \sum_i C_i\left( \frac{M_W}{\mu}\right)  \hat O_i(\mu)
\ ,\eeq
where $\hat O_i(\mu)$ are composite operators renormalized at the
scale $\mu$  and $ C_i\left( {M_W}/{\mu}\right)$ the corresponding
Wilson coefficients. The dependence on $\mu$ of the operators is compensated
by the dependence on $\mu$ of the Wilson coefficients. 
The factorized matrix elements of the operators, however, are 
$\mu$-independent. Thus the final physical amplitude, in the factorization  
approximation, seems to depend on the renormalization scale. It may
be true that ther exists a scale at which non-factorizable contributions
are small, e.g. $\mu=m_b$ for $B$-decays, but a  violation
of factorization must be  present to some extent. 
\par iii) As recently shown in 
ref.~\cite{ley} factorization and quark-hadron duality are incompatible.
To reconcile duality and factorization a certain amount of non-factorizable
contributions has to be allowed. \par
\begin{figure}   
    \centering
    \epsfxsize=0.50\textwidth
    \leavevmode\epsffile{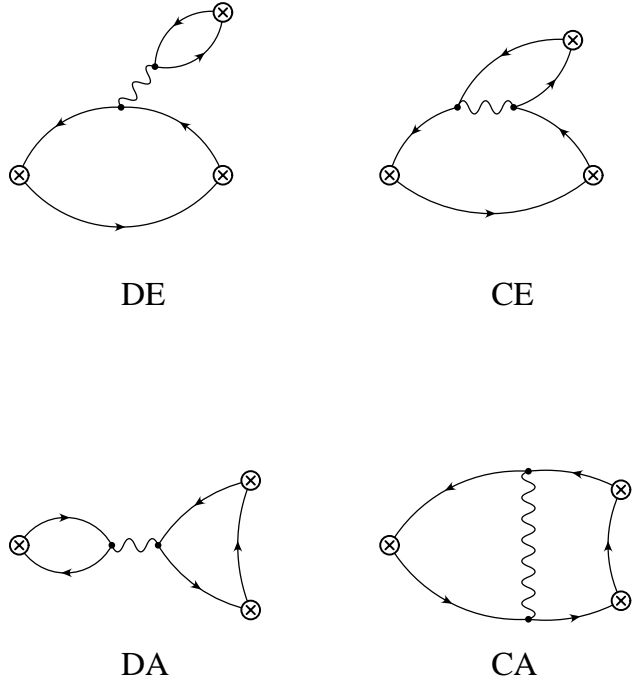}
       \caption[]{\it{Non-penguin Feynman diagrams for
non-leptonic two-body decays.}}
    \protect\label{fig:diags}
\end{figure}

 In conclusion, we know that 
non-factorizable contributions ought to be there, and that they are important
for charm decays. The basic question is then to have a quantitative
estimate of their size in $B$-decays. This problem has been studied
in  extensive experimental  analyses of two-body exclusive 
decays  \cite{artuso} and  it could be possibly accessible to
lattice simulations in the near future \cite{ciu1}, see below.
\vskip 0.2cm 
\underline{Non-leptonic decays on the lattice}:
 There exist only  quite old (and not very accurate) studies of 
heavy-quark two-body non-leptonic decays on the lattice,
performed at $m_Q \sim m_{charm}$ \cite{latnl}.   In these calculations
only ``non-penguin" amplitudes were considered, namely
those   corresponding to
the following Feynman diagrams: disconnected emission (DE),
denoted  also as $T$ or $T^\prime$ in ref.~\cite{gronau1}; the colour-suppressed
connected (non-factorizable)
emission (CE), denoted also as $C$ or $C^\prime$;
the helicity suppressed disconnected annihilation (DA),
denoted also as $A$ or $A^\prime$, and connected annihilation
(CA), denoted also as $E$ or $E^\prime$. These
diagrams are given in  fig.~\ref{fig:diags}.

 In  \cite{latnl} it was  found  that the amplitude
of the  non-factorizable diagram CE was
about one third of the leading contribution given by DE, as expected
on the basis the $1/N$ expansion, and that the amplitude of DA was indeed
compatible with zero. The amplitude
corresponding to CA, which is  assumed to be
negligible in some theoretical analyses \cite{gronau1}, was found,
however, to be as large as one half of that from DE. All the above 
results were  unfortunately afflicted by 
large statistical and systematical errors, and  have never been repeated
since. The reason for this is that all lattice calculations
of two-body non-leptonic amplitudes  
were abruptly interrupted when the Maiani-Testa no-go theorem was 
published \cite{mt}. This theorem states the following.
In numerical simulations on the lattice, it is unavoidable to  
 make a Wick rotation to the Euclidean space-time, where
 $in$- and $out$-states are indistinguishable. From the
standard   study of the correlation functions one  obtains then
the  average of the
$in$- and $out$- amplitudes instead of the physical one, thus loosing
all information about final state interactions~\footnote{In 
ref.~\cite{bloks},
a similar problem was found in the context of QCD sum rules.}.
 For example, in the
calculation of the decay of a $B$-meson into two light mesons $M_1$
and $M_2$, mediated by the weak Hamiltonian ${\cal H}^W$, one 
gets 
\beq A^W= \frac{1}{2} \Bigl(\ _{out}\langle M_1 \ M_2 \vert {\cal H}^W\vert B
 \rangle
+ _{in}\langle M_1\ M_2 \vert {\cal H}^W\vert B \rangle \Bigr) \ ,\eeq
instead of $_{out}\langle M_1 \ M_2 \vert {\cal H}^W\vert B
 \rangle$ only.
Moreover, at large time-distances, the correlation functions 
from which the amplitudes are usually extracted
are dominated
by the off-shell matrix element $\langle M_1(\vec p_1=0) \
 M_2(\vec p_2=0) \vert {\cal H}^W\vert B \rangle$, instead of the 
physical amplitude. In general,
the Maiani-Testa theorem prevents 
the calculation of time-like electro-magnetic form factors and  strong and 
weak decays in two or more hadron final states. 
Its extension can be 
used to show that it also limits the possibility of computing inclusive
cross-sections and decay rates \cite{martipan96}. 
A possible way out of this problem,
under  mild assumptions, although at the price of introducing some
model dependence, has been recently suggested in ref.~\cite{ciu1}.
We do not know at present if the proposal of ref.~\cite{ciu1}
will, in practice, succeed in extracting the physical amplitudes including the
phases due to final state interactions. Feasibility studies are currently 
underway. If the proposal works, a new field of investigation
(factorization, scaling laws, effects due to nearby resonances) will
be open to lattice calculations.
\par \underline{Other interesting two-body amplitudes}:
\par i) Penguin Pollution: in $B^0 \to \pi^+ \pi^-$ decays, CP violating
effects are given by the combined contribution
of  tree-level diagrams and
penguin diagrams~\cite{gronau}.  In order to extract
the relevant phases, it is the necessary to discriminate the two 
effects. Experimentally, this can be achieved
 by separating the term proportional to 
$\cos \Delta M_{B_d} t$ from that proportional to
$\sin \Delta M_{B_d} t$
in the time-dependent asymmetry 
\beq {\cal A} =\frac{N(B^0_d \to \pi^+\pi^-)(t)- N(\bar B^0_d \to \pi^+\pi^-)
(t)}{N(B^0_d \to \pi^+\pi^-)(t)+ N(\bar B^0_d \to \pi^+\pi^-)
(t)} \, .\eeq
A theoretical 
calculation of the tree-level and penguin contributions
 using the method of ref.~\cite{ciu1}, combined with the experimental analysis,
would certainly help to reduce the systematic uncertanties in the final 
results.
\par ii) Isospin analysis: by combining the measured rates of
the $B^0 \to \pi^+ \pi^-$ and $B^0 \to \pi^0 \pi^0$ decays, one
can remove the penguin contribution, by isolating the pure $\Delta I=3/2$
amplitude~\cite{gronau}. The difficulty of this method is due to the measurement
of the decay rate into two neutral pions, which is expected to be colour
suppressed. An alternative method, based
on a combined study of $B^0 \to \pi \pi$, $B^0 \to  \pi K$ and
$B^0 \to K \bar K$ decays, using approximate flavour $SU(3)$ symmetry
has also been proposed. There is no space
to  discuss   the details, and uncertainties,
of this method, for which the reader can refer to~\cite{gronau}.
I want only to add one observation, which may stimulate further investigation.
In refs.~\cite{gronau1,gronau}, it is usually assumed that the non-factorizable
annihilation diagram CA is negligible, in contrast, but
within large errors, and for smaller heavy quark masses,
with the results of ref.~\cite{latnl}, see above. 
A reappraisal of the lattice calculations of the non-leptonic 
amplitudes, on large lattices,
and by using the theoretical improvements developed in the  last few years
\cite{martipan96} would certainly be very useful in this respect.
\section{Power corrections in the HQET}
\label{sec:power}
In this section the problem of controlling power corrections
 in the heavy quark effective theory (HQET) is discussed. 
These corrections are necessary
for an accurate  evaluation of both inclusive and exclusive decays of hadrons 
containing a $b$-quark.
The basic question is whether they can be computed
 with a theoretical error smaller than the corrections themselves
\cite{msreno}.
\par  To be specific,  we start by considering  the generic expression 
for the inclusive decay rate \cite{cgg}--\cite{ns}
\beqn \Gamma( H_b \to X_f)= \frac{G_F^2 m_b^5}{192 \pi^3} \left\{
\left(1+ \frac{\lambda_1 + 3 \lambda_2}{2 m_b^2}\right) \ C^3_f
+ \frac{6 \lambda_2}{m_b^2} \ C^5_f + \frac{16 \pi^2 f_B^2 M_B}{m_b^3}
\sum_i C_i^6 B_i^6 + \dots \right\} \ , \label{eq:inclu} \eeqn
and the ratio of form factors in  $B \to D^* l \nu_l$ decays \cite{hqet}
\beqn \frac{h^V(\omega=v \cdot v^\prime)}{h^A(\omega=v \cdot v^\prime)}
= \frac{C^V(\omega)}{C^A(\omega)} \left( 1 + \frac{\labar}{m_c} f(\omega)
+ \dots \right) \ . \label{eq:exclu} \eeqn
In the above expressions $h^{V,A}$ are the relevant vector
and axial vector form factors;  $C^3_f$, $C^5_f$, $C_i^6$,
$C^V(\omega)$ and $C^A(\omega)$ are short distance quantities that can be 
computed in perturbation theory; the quark mass $m_b$, 
the binding energy $\labar$, the kinetic energy $\lambda_1$, the 
chromo-magnetic term
$\lambda_2$ and the $B^6_i$'s, that are related to the matrix elements
of four-fermion operators, require instead a non-perturbative evaluation. 
\par In some cases the non-perturbative parameters can be directly related,
at the leading order in $1/m_Q$, to a physical quantity.
 This is the case  of the chromo-magnetic term which corresponds to the $B^*$--$B$
mass splitting, $\lambda_2=1/4 (M_{B^*}^2-M_B^2)$.  In general, however,
 the coefficients of the $1/m_Q$
expansion, and the corresponding non-perturbative parameters, are plagued
by renormalon ambiguities. This problem  is  present for example  
in  the definition  the binding energy $\lbar$ appearing in eq.~(\ref{eq:exclu})
 and in the matrix elements $ B_i^6$
of the four fermion operators in eq.~(\ref{eq:inclu})~\cite{msreno},
\cite{rus}--\cite{bbz}.
In these cases, the cancellation of the ambiguities occurs between
the matrix elements of the higher dimensional operators and the Wilson
coefficient functions ($C^3_f$, $C^5_f$, $...$).
The latter have to be computed  to sufficiently high order of 
perturbation theory for the cancellation to be under
control. Let us imagine that we want to  achieve an
accuracy of $O(1/m_b^k)$. A simple argument indicates 
that, in principle, the leading Wilson coefficient has to be computed
at least to an order such that 
$\alpha_s(m_b)^n \sim (\Lambda_{QCD}/m_b)^k$~\cite{msreno,mueller}.
 Since, in the calculations performed up to
now, only the first few terms of the perturbation series are known, it
is not possible to check that the remaining terms are 
negligible. By studying some simple examples, it has been shown that the
knowledge of only a few terms is, in general, insufficient to control
the power corrections \cite{msreno}. 
Although the results 
 rely on some approximations that were adopted in these examples, it is
likely that the  conclusions will remain valid in general, and that in
most cases a further theoretical effort is needed.
\par Let us consider, for example, the unphysical parameter $\labar=
M_H-m_{pole}$ which is frequently used in phenomenological 
studies of $B$-physics.  Results and bounds for $\labar$ have been presented
in the literature \cite{varie}--\cite{grem}. In order
for this to make any sense, $\labar$ must be defined precisely in terms 
of some physical quantity, and its value will depend on this physical quantity
and  on the order of perturbation theory used to extract $\labar$.  In
 ref.~\cite{msreno} it has been shown that, even by using
a consistent and renormalon-free definition of the binding energy,
 one expects that the value
 of $\labar$  will change by several hundred MeV, i.e. by an amount
comparable to its value, as the order of
 perturbation theory is increased. The error on $\labar$ can be seen
as an error on the $\MSbar$ mass  of the $b$-quark $m_b^{\msbar}(m_b^{\msbar})$
as derived from the HQET \cite{cgms}
\beq m_b^{\msbar}(m_b^{\msbar})= 4.17 \pm 0.05 \pm 0.20 \, \mbox{GeV} \ , \eeq
where the last (and larger) error comes from an estimate of the uncertainty due
to higher order corrections in perturbation theory. \par
 Similar problems are encountered in the case of $\lambda_1$.
One example is given by the zero
recoil sum rules \cite{bounds,bounds2}. In this case,
 in order to derive  bounds on the
relevant form factors, and on   the heavy-quark
kinetic energy parameter $\lambda_1$, a physical cut-off 
$\Delta$ is introduced,  to suppress the contribution
of states with excitation energies greater than $\Delta$. The difficulty
in achieving an accurate determination or bound with a physical cut-off
is due to the higher order perturbative corrections and only by 
arriving to an order such that $\alpha_s^n \le O(\lambda_1/\Delta^2)$
it will be possible to obtain significant results.
A connected example is provided by the different theoretical
determinations of $\lambda_1$, which are given in table \ref{tab:l1}.
Some of them are not compatible, if we accept as realistic the
errors quoted in the original papers. I believe that the differences
are due to higher order perturbative corrections, which have not been
computed, or estimated, yet.
\begin{table} \centering
\begin{tabular}{||c|c|c||}
\hline
\hline
\multicolumn{3}{||c||}{Theoretical estimates of $\lambda_1$ }\\
\hline \hline
Reference &
\multicolumn{1}{c|}{Method}&
\multicolumn{1}{c|}{$-\lambda_1$ (GeV$^2$) }\\
\hline \hline
Eletsky and Shuryak \cite{es} &  QCD sum rules & $0.18 \pm 0.16$ \\
Ball and Braun \cite{qcdsr}& QCD sum rules &$0.52\pm 0.12$  \\
Bigi et al. \cite{bounds} & ZR sum rules &$\ge 0.36$  \\
Kapustin et al. \cite{bounds2}
& ZR sum rules +$O(\alpha_s)+O(\alpha_s^2 \beta_0)$
&$--$  \\
Ligeti and Nir  \cite{zn}& Experiment&$\le 0.63$ if $\labar \ge 240$ MeV  \\
 & Experiment&$\le 0.10$ if $\labar \ge 500$ MeV  \\
Gim\'enez et al. \cite{newl1}& Lattice & $-0.09 \pm 0.14$ \\ 
Chernyak  \cite{cher} & Experiment& $0.14\pm 0.03$  \\
Gremm et al. \cite{grem}& Experiment&$0.19\pm 0.10$  \\
Neubert \cite{virial} & Virial Theorem & $0.10 \pm 0.05$ \\
\hline\hline
\end{tabular}
\caption{\it{Some of the 
values of  $\lambda_1$ obtained in different theoretical
analyses. ``Experiment" denotes the extraction of $\lambda_1$ from the
experimental data, for example the charged-lepton spectrum distribution
in semi-leptonic $B$-meson decays. ``ZR sum rules" denotes the zero recoil
sum rules.}}\label{tab:l1}
\end{table}
\section{Semi-leptonic inclusive decays} 
\label{sec:insemilepto}
The idea that inclusive decay rates of heavy quarks can be computed 
in the parton model is quite old and was used, for example,
 in ref.~\cite{cabma} to predict charm hadron lifetimes. To account
for bound state effects, the partonic  calculation was 
successively improved by the introduction of a
  phenomenological model, called  
the ``spectator model"  \cite{pieta,accmm}, which has
been and continues to be
 extensively used to extract $\vert V_{cb}\vert $ and $\vert V_{ub}
\vert$ from inclusive semi-leptonic decays. A noble theoretical framework
for the spectator model was then provided  by the operator product expansion
applied to inclusive decays of heavy quarks \cite{cgg}--\cite{ns}.
The general result is summarized in eq.~(\ref{eq:inclu}), which holds 
for both inclusive semi-leptonic and non-leptonic decays.
The first term on the r.h.s. of eq.~(\ref{eq:inclu}) is the parton model
 result;
 corrections of $O(1/m_b^2)$ are due to the kinetic energy $-\lambda_1$
and the spin interaction $\lambda_2$ of the heavy quark in the 
hadron~\footnote{ Up to a redefinition
of the quark mass, there is indeed a relation between the Fermi momentum
of the heavy quark in the model of ref.~\cite{accmm} and 
$\lambda_1$.};
 the four-fermion operator  contributions to the $1/m_b^3$ correction,
written in terms of the $B^6_i$ parameters, correspond, in the old
language, to   W-exchange and interference higher-twist effects.
These effects  were considered in order to explain the lifetime differences
among the different charmed hadrons \cite{guberina}. 
\par In modern language,
the use of the parton model to compute inclusive decays relies on 
the assumption of  ``quark-hadron" duality: 
in semi-leptonic decays, where the integration over 
the lepton and neutrino momenta corresponds to an integration over the
 invariant mass of the recoiling hadronic system, we have global
duality, whereas in non-leptonic
 decays, where the hadronic mass is fixed to be equal to the heavy hadron
mass, we  speak of local duality. Duality is assumed, but cannot be
proven in QCD. We know that it provides a good description of several
physical processes such as inclusive deep inelastic scattering,
inclusive $e^+ e^-$ annihilation into hadrons, Drell-Yan processes, etc.
The assumption of global duality seems also to work for semi-leptonic
(and radiative)  decays. This means that we are able to predict
the inclusive $\Gamma_{SL}(b \to X_c l \nu_l)$,
$\Gamma_{SL}(b \to X_u l \nu_l)$ and  $\Gamma_{SL}(b \to s \gamma)$
rates. This is not
the case  for local duality since  there is  strong experimental
evidence that we are unable to explain $B$-hadron lifetimes in the framework
of the $1/m_b$ expansion, see sec.~\ref{sec:innonlepto}.
\par The situation for $\Gamma_{SL}(b \to X_c l \nu_l)$ is substantially
unchanged since last year. For this reason, I direct the reader to 
ref.~\cite{nrev} for a discussion of the theoretical uncertainties
in the calculation of this quantity. Here I limit myself giving the
upgraded world average values of  $\vert V_{cb}\vert $ from
inclusive decays
\beqn \vert V_{cb}\vert^{\Upsilon(4s)} &=& 0.0391 \pm 0.0007 \pm 0.0030 \ ,
\nonumber \\
 \vert V_{cb}\vert^{LEP} &=& 0.0412 \pm 0.0004 \pm 0.0030
\label{eq:vcbin}\ ,\eeqn
where the results from low energy and LEP energies have been given separated.
The reason is that the different central values are induced mostly by
 the  differences in the semi-leptonic branching ratio  measured
in the two cases \cite{richman}. Notice the excellent agreement 
between these results and the number given in eq.~(\ref{eq:vcbex}).
\par The above results make us confident that we are able to make
accurate predictions for $\Gamma_{SL}(b \to X_c l \nu_l)$. A related problem
is whether we are also able to predict the semi-leptonic branching ratio
$B_{SL}$, and the average number  of charmed hadrons $n_c$.
Since the former is given by the ratio of  $\Gamma_{SL}(b \to X_c l \nu_l)$
to the total width, the problem is whether we are able to predict
inclusive non-leptonic widths. Recently, there  was a lot of debate 
about whether
the theoretical value of $B_{SL}$ is in agreement, or in disagreement,
with the experimental measurements
\cite{ns,bigi2,bagan}  and, in the second case, several
explanations, within and beyond the standard model, have been proposed
\cite{ampr}--\cite{giu}. Since there are important differences
between low energy and LEP results \cite{richman}
\beqn B_{SL}^{\Upsilon(2s)}&=& 10.49 \pm 0.17 \pm 0.43 \ ,
\;\;\;\;\;\;\;\;\;\; n_c^{\Upsilon(2s)}= 1.12 \pm 0.05 \nonumber \\
B_{SL}^{LEP}&=& 10.95 \pm 0.13 \pm 0.29 \ ,
\;\;\;\;\;\;\;\;\;\; n_c^{\Upsilon(2s)}= 1.22 \pm 0.08  \eeqn
it is probably premature to draw any conclusion at this stage.
\par  $\vert V_{ub} \vert$ could be extracted, in principle, from the
total semi-leptonic width in the same way as $\vert V_{cb} \vert$.
In practice, however, only the endpoint spectrum is used, in order to
eliminate the background due to $ b \to c  l \nu_l$ decays.
The cut on the lepton energy $E_l$ is fixed by
the requirement that the recoiling
hadronic system $X$ has an invariant mass $M_X \le M_D$, so as to 
inhibit charm decays. In this region, however, we run into troubles
with the operator product expansion. Schematically,  we can write
\beq \frac{d\Gamma(b \to u l \nu_l)}{dE_l}=
\sum_n \int_{y_{min}(E_l)}^{y_{max}(E_l)} dy \left(\frac{\Lambda_{QCD}}{m_b
(1-y)}\right)^n \Gamma_n(y) + \dots\ , \label{eq:vubin} \eeq
where $\dots$ represent terms which are suppressed as powers of $m_b$.
Only the first few terms in the expansion above are known 
because they can be related to $\lambda_1$, $\lambda_2$,  etc. The  higher
order terms depend on  matrix elements of higher dimensional
local operators, which can at most be estimated within a specific
phenomenological model. \par 
  Near the end point, $y_{min} \sim 1-M_X^2/m_b^2$ ($y_{max}$
is always of $O(1)$)  so that when $M_X \sim
\sqrt{\Lambda_{QCD} M_B}$, all the terms in the expansion (\ref{eq:vubin})
become equally important. At even larger values
of $E_l$, $M_X \sim \Lambda_{QCD}$. In this case we are in the elastic
region (corresponding to $B \to \rho$, $B \to \pi$, etc.) and the decay
cannot be analyzed with operator product expansion.  
Unfortunately, $M_X \sim \sqrt{\Lambda_{QCD} M_B}$ is about
$1.2$--$1.3$ GeV, a value very close to the  kinematical cut
which is currently used in the experiments  to extract  $\vert V_{ub} \vert$. 
This implies  that the extraction of $\vert V_{ub} \vert$ is still
subject to an intrinsic model dependence, which can be reduced either
by enlarging the region of $E_l$ from where $\vert
V_{ub} \vert$ is extracted, at the price of increasing the 
charm background, or by a sound theoretical prediction of the
``wave" function \cite{bigi,mannel} of the heavy quark.
\section{Non-leptonic inclusive decays} 
\label{sec:innonlepto}
In this section it will be shown that it is very unlikely that 
higher order   corrections  in $1/m_b$ can explain the discrepancy between the
theoretical predictions  for the $B$ to $\Lambda_b$ lifetime ratio
$R=\tau_{B}/\tau_{\Lambda_b}$. 
\par
 For $R$, the dependence on the
fifth power of the quark mass, as well as all  renormalon ambiguities
up to order $1/m_b^3$, cancel out. We thus expect  $R$ to be one of the 
soundest predictions of  HQET.  The discrepancy between theory and
 experimental data  is evidence for the failure of local
duality  in  the calculation of the non-leptonic (NL) widths of heavy hadrons. 
\par
To order $1/m_b^2$ in the heavy quark expansion, the lifetime ratio $R$
is given by
\beq R= 1 + \frac{\lambda_1(B)-\lambda_1(\Lambda_b)}{2 m_b^2}+
C_G \frac{\lambda_2(B)-\lambda_2(\Lambda_b)}{ m_b^2}+O(1/m_b^3) 
\label{ratio} \ , \eeq
where the difference of kinetic energies $\lambda_1(B)-\lambda_1(\Lambda_b)
= 0.01 \pm 0.03$ GeV$^2$ can be obtained from the spectrum of the
heavy hadrons; $\lambda_2(\Lambda_b)=0$; $\lambda_2(B)=(M_{B^*}^2-
M_B^2)/4=0.12$ GeV$^2$ and $C_G/3 \sim 1.1$. Using the numerical values 
given above one obtains $R=1.02 +O(1/m_b^3)$, incompatible with the 
observed ratio of lifetimes $R=1.27 \pm 0.05$. Within the standard
HQET, in order to explain the large deviation from unity for this ratio it 
is necessary the the matrix elements of the four-quark operators entering
in the $1/m_b^3$ corrections (and in particular those relative to the
$\Lambda_b$) be  much larger than those estimated in quark models, and
more recently with QCD sum rules \cite{ns,cola}. 
In my opinion, this remains an open, but 
very unlikely possibility.  An alternative explanation is that for NL
widths there is evidence, contrary to the orthodox expectations, of corrections
of $O(1/m_b)$. These corrections are well described  by the simple ansatz 
that replaces the
quark  mass  with the decaying  hadron mass  in  the $m_Q^5$ factor in front
of the NL width \cite{martihf95,ampr}
\beqn \Gamma( H_b \to X_f)= \frac{G_F^2 m_b^5}{192 \pi^3}\vert_{HQET} 
+ O(1/m_b^2)
\to  \frac{G_F^2 M_H^5}{192 \pi^3} + O(1/M_H^2) \ .
\label{eq:3} \eeqn
This replacement provides a much better description of the NL widths. 
In ref.~\cite{ampr}, it was
shown that, for beauty, both the problems of the SL branching ratio and of
the difference in the lifetimes of the $\Lambda_b$ baryon and the $B$ mesons
are quantitatively solved. For charm a much better fit to the seven known
lifetimes is obtained in terms of four parameters of reasonable size: one
lifetime, one interference contribution for $D^+$, one for $\Xi^+$ and a
smaller W-exchange term for $D^s$. 
I will now give  more details on the analysis of ref.~\cite{ampr}. 
\par 
Given the agreement between
the value of $\vert V_{cb} \vert$  extracted from inclusive $b \to c$
and exclusive $B \to D^*$ semi-leptonic (SL) decays and the approximate
equality of the SL widths for charmed hadrons (within large uncertainties
and with a rather limited number of cases where  measurements exist), it is
 assumed that the HQET  is able to predict the SL widths
$\Gamma_{SL}$ of the heavy hadrons. Under this assumption,
 and using the ansatz (\ref{eq:3}), one gets
 \beq R  = ({m_{\Lambda_b}\over
m_B})^5 \left[ 1-2.24 B_{SL} (B)\right] + 2.24 B_{SL} (B) +
O(1/M_H^2) \ , \label{eq:4} \eeq
 where the factor 2.24 arises from taking
the electron, the muon and the tau SL modes in the ratio 1:1:0.24
\cite{Neub}, and the difference in the SL rates of the  $\Lambda_b$ and $B$
has been neglected. From $M_{\Lambda_b} = 5623 \pm 6$~MeV \cite{bede} and
$M_B = 5279 \pm 2$~MeV \cite{pdg}, by using 
$B_{SL}(B)=10.77 \pm 0.43 \%$, one finds~\footnote{ The same numbers
as in ref.~\cite{ampr} are used here since the analysis has not been
repeated since.} 
\beq{\tau_B\over \tau_{\Lambda_b}} = 1.29 \pm 0.05\label{eq:5} \eeq
in perfect agreement with the experimental result.  

So far this agreement can
be considered to be  a  numerical accident, lacking  theoretical explanation. 
For this reason, it is  quite interesting to consider what happens in the case of charmed
particles for which seven lifetimes have been measured, see table \ref{tab:lt}.
\begin{table} \centering \begin{tabular} {||c|c|c|c|c|}\hline Hadron    & 
Mass  (MeV/$c^2$) & $ \tau$ (ps)   & $ B_{SL} ($\%$) $ & $ \Gamma_{SL}=
B_{SL}/ \tau$ (ps$^{-1}$) \\ \hline $D^{\pm}      $  & $1869.4 \pm 0.4$ &
$1.057 \pm 0.015$ & $17.2 \pm 1.9$ &$16.3\pm 1.8$\\ $D^{0}        $  &
$1864.6 \pm 0.5$ & $0.415 \pm 0.004$ & $ 8.1 \pm 1.1$ &$19.5\pm 2.6$\\
${D_s}        $  & $1968.5 \pm 0.7$ & $0.467 \pm 0.017$ &                &             
\\ $\Lambda_c^{0}$  & $2285.1 \pm 0.6$ & $0.200 \pm 0.011$ & $ 4.5 \pm 1.7$
&$22.5 \pm 8.5$\\ $\Xi_c^{0}    $  & $2470.3 \pm 1.8$ & $0.098 \pm 0.019$ &               
&              \\ $\Xi_c^{\pm}  $  & $2465.1 \pm 1.6$ & $0.350 \pm 0.055$ &              
&           \\ $\Omega_c^{0} $  & $2704   \pm 4  $ & $0.055 \pm 0.023$ &              
&           \\ \hline \end{tabular} \caption{\it{ Properties of charmed mesons
and baryons; the $\Omega^{0} $ values are an average of the data quoted in
ref.~\protect\cite{wa89}.}}\label{tab:lt} \end{table}
At lowest order in the
$1/m_Q$ expansion,  a much better  agreement with the experimental results 
for the lifetimes is obtained by replacing the heavy-quark mass by the
hadron masses  as in eq.~(\ref{eq:3}).
We neglect at this stage any other mass correction and  write  $\Gamma_{NL}
(M_H) = \Gamma_{tot}(M_H) - 2\Gamma_{SL}$, where for $\Gamma_{SL}$  a
universal value chosen as the average of the experimental values for $D^+$,
$D^0$ and $\Lambda_c$, or  $\Gamma_{SL} = (0.174 \pm 0.015) $ps$^{-1}$
is used \cite{pdg}. The dependence on the hadron mass of $\Gamma_{NL}(M_H)$ will be
taken according to $\Gamma_{NL}(M_H) = (M_H/M_0)^n \Gamma_{NL}(m_0)$ with n = 5,
where $M_0$ is around the average mass of the relevant hadron. We then have
\beq \Gamma_{tot} (M_H) = \tau^{-1} (M_H) = \tau^{-1} (M_0) ({M_H\over M_0})^n +
2\Gamma_{SL} (1 - ({M_H\over M_0})^n)\label{eq:10} \eeq
\begin{figure}   
\begin{center}
\setlength{\unitlength}{1truecm} 
\begin{picture}(7.5,10.) 
\put(-8.0,-9.5)
{\includegraphics{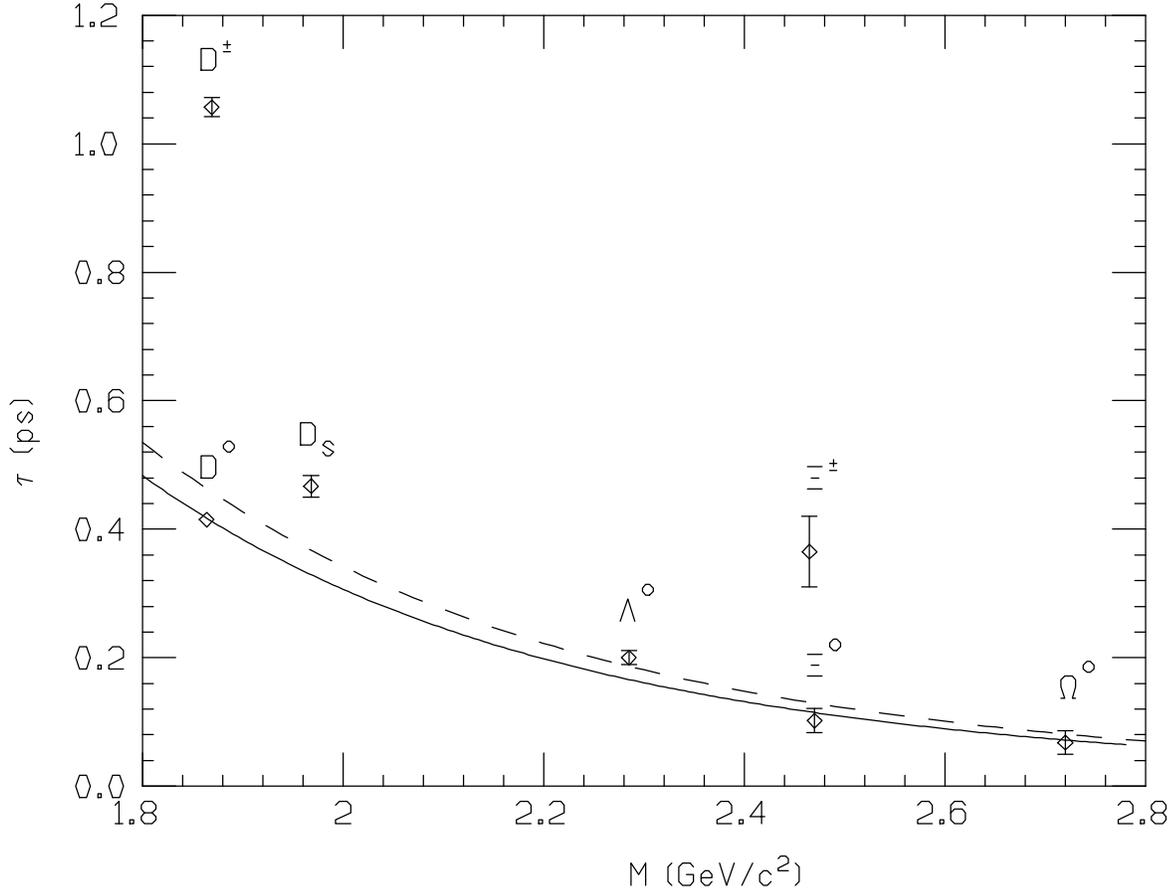}} 
\end{picture} 
\end{center} 
\vskip 1.0cm 
\caption{\it{Lifetime vs. mass for charmed particles. The dashed line is the
best fit described in the text for all seven points assuming proportionality
of the NL widths to $M_{H}^5$, $M_H$ being the hadron mass. The solid line
is the best fit restricted to only the $D^0$, $\Lambda_c$, $\Xi^0$ and
$\Omega_c$ lifetimes with the same assumptions as before.}}
\protect\label{uno}
\end{figure}
 
Let us fix $n=5, M_0 = 2.3$ GeV and $\Gamma_{SL}= 0.174 \, $ ps$^{-1}$ and
fit all seven known lifetimes in terms of  $\tau (M_0)$.  $\tau
(m_0) = 0.181$ ps is then obtained. 
The corresponding fit is shown in fig.~\ref{uno}
 (dashed curve).
We see that four out of seven lifetimes are in very good agreement with the
fitted curve. The lifetimes of $D^+$, of $\Xi^+$  and, to a lesser extent,
of $D_s$ are clearly out. The discrepancies for $D^+$ and
$\Xi^+$ can be attributed to the interference effect \cite{bigiic}. 
Note that $D^+$  is the only
meson that can have interference at the Cabibbo allowed level and $\Xi^+$ is
the only baryon that can have double interference, in the sense that  $\Xi^+
= cus$ and both $u$ and $s$ can interfere with the corresponding quarks from
$c \rightarrow su \overline {d}$. For $D_s$  the observed smaller difference
is attributed to the possibility of W-exchange \cite{bigiic}. All of these
effects are of order $f^2_D /m_c^2$ or $1/m_c^3$.  The solid line in 
fig.~\ref{uno} has
been obtained  from a modified fit where only  the $D^0$, $\Lambda_c$,
$\Xi^0$ and $\Omega_c$ lifetimes have been considered. In this case,
  $\tau (M_0) = 0.161$ ps, with the respectable value of the
$\chi^2/d.o.f.$ given by $\sim 3.5$, is obtained. 
For  comparison, the fit of the quark
mass to constant lifetimes results  in a  $\chi^2/d.o.f.\sim  250$. 
Finally, for the same four lifetimes, one fits the  power $n$ in
eq.~(\ref{eq:10}), keeping fixed the value of  $M_0$  and $\tau(M_0)$ at the
observed values for the $D^0$ meson. In this way one may  check whether the best
power for $n$  is close to $5$.  $n = 4.5 \pm 0.5$ is found, where the error
arises from the experimental errors on the lifetimes. Moreover, if we write
for $D^+$, $\Xi^+$ and $D_s$  the expression $ \tau^{-1} = \tau^{-1} (M_H)
\left[1 - ({\mu\over M_H})^3\right]$, where $\tau$ is the experimental
number given in table~\ref{tab:lt} and  $\tau(M_H)$ is taken from the previous fit  to
the four remaining lifetimes (with n=5 ), $\mu = 1.6, 2.2$ and
$1.3$~GeV are obtained for $D^+, \Xi^+$ and $D_s$, respectively. We
 see that the resulting values of this correction 
are large, as it is obvious from  fig.~\ref{uno}, but not unreasonable.
\par
The conclusion is that 
a number of experimental facts show that  $\Gamma_{NL}$ for charm and beauty
 decay approximately
scale with the fifth power of hadron masses (apart from corrections of order
$1/m_Q^2$ or smaller) rather than having $O(1/m_Q^2)$ corrections
to the universal behaviour predicted by the HQET. Since $M^5_{H_b}=
(m_b +\bar \Lambda_{H_b})^5 = m_b^5 (1 + 5 \Lambda_{H_b}/m_b) +\dots$,
  this  implies that the corrections are of $O(1/m_b)$.
Since the validity of the
operator expansion in the timelike region, in the vicinity of the physical
cut, is not at all guaranteed \cite{pqw,shifm},   the
failure of the short distance approach can be attributed
to a violation of the local duality
property that has to be assumed for NL widths. This is to be
compared to the case of the  SL widths for
which the predictions of the HQET are  not inconsistent with the data.
The experimental evidence for NL widths calls for a reexamination of the underlying
theoretical framework. An attempt in this direction can be
found in ref.~\cite{uralnl}.
 More extended and accurate measurements of the
SL widths of each hadron separately are also needed in order
 to assess the validity of the heavy quark expansion in semi-leptonic decays. 
\section{Conclusion and outlook}
\label{sec:conclu}
Beauty hadrons  provide an extraordinary laboratory 
to test the Standard Model, to look for hint of new physics and
to understand QCD dynamics. For this reason an exceptional
experimental and theoretical effort has been invested in it.
In this review, I have tried to point out those dynamical  aspects
which we ought to understand and control at a quantitative level
in order to extract the relevant information on some
fundamental parameters, such as  the mass of the $b$ quark,
the relevant CKM matrix elements and CP violation. 
\par Much theoretical progress, and improvement in understanding, has
occured this year in the framework of the HQET and of the
calculation of inclusive rates. Still a lot of work remains to 
be done, particularly for those quantities where the theoretical 
uncertainties are larger than the experimental errors.
Progress can be expected quite soon for the slope and the curvature
of the form factors in $B \to D^* l \nu_l$ decays, in the understanding
of factorization and in the calculation of non-leptonic 
two-body decay amplitudes. It appears more difficult to reduce 
the error in the normalization of the form factor at zero recoil
for $B \to D^* l \nu_l$ and to eliminate the model dependence in
the extraction of $\vert V_{ub} \vert$ from inclusive
semi-leptonic decays. A real puzzle, which demands a deep reappraisal
of current ideas, and prejudices, remains that of the $\Lambda_b$
lifetime, which seems to indicate a failure of local duality
in  inclusive non-leptonic decays.
\section*{Acknowlegments}
I thank M.~Artuso, A.J.~Buras, J.~Flynn, M.~Gronau, A.~Masiero,
M.~Neubert, C.T.~Sachrajda, S.~Sharpe,
and N.~Uraltsev for informative, interesting, and intense discussions.
I acknowledge the partial support by M.U.R.S.T.  and
by the EC contract CHRX-CT92-0051.
    
\end{document}